# CARACTERISATION ELECTROMAGNETIQUE IN-SITU DE SOLS EN BANDE L. APPLICATION A L'IDENTIFICATION DE PROFILS GEOLOGIQUES.

[1]F. Demontoux, [1]G. Ruffié et [1]D. Medina Hernandez

[1]Université Bordeaux 1 – Laboratoire IMS UMR 5218 – département MCM – 16 avenue Pey-Berland 33607 Pessac - France

francois.demontoux@ims-bordeaux.fr ; gilles.ruffie@ims-bordeaux.fr

**Résumé :** Le développement des instruments de mesure des applications de télédétection spatiale (active ou passive) nécessite des phases de validation in-situ. Les équipements (radar ou radiomètre) doivent être testés sur des profils géologiques réels. Nous devons donc posséder des informations précises sur ces profils étudiés (permittivités, humidité, épaisseurs de couches…). Des capteurs (humidité), des carottages (épaisseurs) ou des mesures en laboratoire (permittivité) nous donnent des informations ponctuelles sur le profil considéré alors que les équipements utilisés en télédétection effectuent une mesure globale sur un volume plus ou moins important de la structure géologique. Les travaux que nous présentons décrivent le banc de caractérisation in-situ, basé sur deux antennes cornets, associés à un logiciel d'identification de profils que nous avons développé pour obtenir des informations globales et instantanées sur les profils géologiques.

**Mots clés:** Bancs de caractérisation in-situ, identification de profils géologiques, télédétection.

## 1. Introduction

Le développement des instruments de mesure des applications de télédétection spatiale (active ou passive) nécessite des phases de validation in-situ. Les équipements (radar ou radiomètre) doivent être testés sur des profils géologiques réels. Nous devons donc posséder des informations précises sur ces profils étudiés (permittivités, humidité, épaisseurs de couches…). Des capteurs (humidité), des carottages (épaisseurs) ou des mesures en laboratoire (permittivité) nous donnent des informations ponctuelles sur le profil considéré alors que les équipements utilisés effectuent une mesure globale sur un volume plus ou moins important de la structure géologique. Pour obtenir des informations globales et instantanées sur le profil nous avons mis au point un banc de caractérisation in-situ, basé sur deux antennes cornets, associé à un logiciel d'identification de profils.

Le travail que nous présentons décrit notre banc de mesure et le logiciel d'identification qui lui est associé. Le principe de ce système d'exploitation repose sur l'utilisation d'une base de données renfermant les réponses électromagnétiques de nombreuses configurations géologiques. Un utilitaire, programmé sous Matlab, permet de comparer les mesures obtenues par le banc avec des résultats simulés. Ces derniers sont générés en utilisant le logiciel HFSS (High Frequency Structure Simulator) de la société ANSOFT [1]. Ceci rend possible la caractérisation électromagnétique des milieux considérés ainsi que la définition de caractéristiques morphologiques, telle que l'épaisseur des différents couches qui composent un milieu.

## 2. Description des outils utilisés

### 2.1 Banc de mesure in-situ et modèle numérique

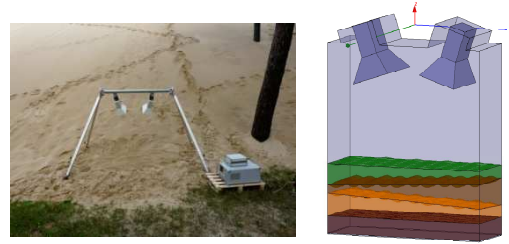

**Figure 1.** *Banc de mesure et modèle numérique.*

Le banc de mesure in-situ est constitué d'un portique, de deux antennes cornets [2] et d'un analyseur de réseau vectoriel portable ANRITSU 2026 A (figure 1). Les antennes sont choisies en fonction de la gamme de fréquence de mesures désirées. Le modèle numérique représentant le banc de mesure a été créé à l'aide du logiciel HFSS et est basé sur la méthode des éléments finis pour la résolution des équations de Maxwell.

### 2.2 Principe du logiciel d'identification de profils

Le système d'exploitation va comparer la courbe expérimentale, nommée Cexp, avec une base de données de courbes simulées, nommées Ci. Le système choisira plusieurs courbes optimales, en fonction de différents critères implémentés. Cette structure est représentée dans la figure 2.

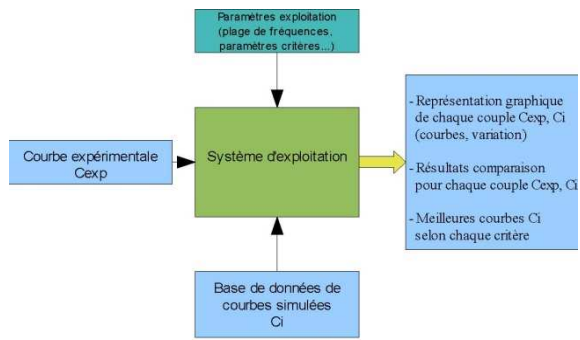

**Figure 2 :** *Structure du logiciel d'identification*

L'objectif n'est pas une identification automatique des caractéristiques électromagnétiques du milieu mesuré. Le système permettra l'exploitation des mesures en réflexion ou en mesure bi-statique des milieux géologiques, facilitant la caractérisation des milieux.

La démarche pour l'exploitation passe par une modélisation du système de mesure et du milieu mesuré. Une fois la modélisation faite, il est possible d'obtenir une base de données de courbes en fonction de plusieurs paramètres du milieu. Ces paramètres pourront être les caractéristiques électromagnétiques (permittivité), ou des caractéristiques morphologiques du milieu, tel que l'épaisseur d'une couche de matériau ou la rugosité [3].

Il est aussi possible d'envisager la construction d'une base de données contenant des courbes générées pour plusieurs configurations de milieu (monocouche, bicouche, avec rugosité...) et pour plusieurs épaisseurs et permittivités.

Si la base de données ne contient aucun des cas équivalents au milieu étudié, une nouvelle modélisation est nécessaire.

Le système va suggérer une ou plusieurs courbes optimales. À la fin, ce sera à l'utilisateur de décider quelles courbes sont des solutions possibles, en prenant en compte des paramètres des courbes simulées. Par exemple, une courbe correspondant à un milieu monocouche d'une certaine épaisseur et permittivité peut avoir la même forme que la courbe d'un système bicouche avec des épaisseurs différentes et des permittivités différentes sur une plage limitée de fréquences. Plusieurs solutions sont alors possibles, et c'est l'utilisateur, qui, en connaissant les conditions de la mesure ou en changeant la largeur de la plage de fréquence pourra lever l'indétermination et pourra choisir finalement la meilleure courbe.

## 2.3 Structure interne du système d'exploitation programmé avec Matlab

Les données d'entrées au système sont les noms des fichiers de mesure et ceux simulés et les paramètres nécessaires pour définir les comparaisons. La courbe expérimentale est contenue dans un fichier texte généré par l'analyseur de réseaux ANRITSU 2026 A utilisé. Les données simulées sont aussi enregistrés dans des fichiers texte générés par le logiciel HFSS. Dans un premier temps, le programme lit ces fichiers et garde en mémoire toutes les données. Puis des comparaisons selon différents critères sont effectuées. Enfin, les résultats des comparaisons sont présentés sous forme de courbes.

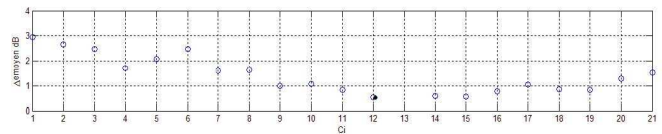

**Figure 3 :** *Courbe de présentation de critère de comparaison*

Pour chaque critère, une série de points présente le résultat obtenu pour chaque comparaison d'une courbe simulée $C_i$ avec la mesure (figure 3). Un point noir indique la meilleure courbe pour chaque critère. Un point rouge indique l'impossibilité d'appliquer le critère pour la courbe $C_i$ donnée.

Plusieurs critères ont été implémentés. Ces critères ont été choisis en tenant compte les caractéristiques des courbes mesurées.

Le premier critère est le décalage entre les pics de résonance. Le programme détecte les pics (minimums) présents dans la courbe expérimentale et les courbes simulées et calcule les décalages en fréquence entre eux. Pour chaque pic, la meilleure courbe sera celle avec un décalage minimal. Pour ce premier critère, plusieurs courbes de résultats seront présentées si plusieurs pics sont présents. Le programme recherche les pics de la courbe expérimentale et les pics des courbes simulées. Si le nombre de pics n'est pas le même, un point rouge apparaît dans la courbe de résultats pour la courbe $C_i$ correspondante. Si le nombre de pics est le même, les décalages sont calculés et présentés sur une courbe spécifique à chaque pic. Il faut noter que la détection des pics de résonance équivaut à la détection des minimums plus prononcés dans les courbes. L'algorithme développé utilise un filtrage passe-bas pour cette détection. C'est la fréquence de coupure de ce filtre qui va déterminer les pics détectés. Il est donc possible de changer cette fréquence pour d'autres types de courbes.

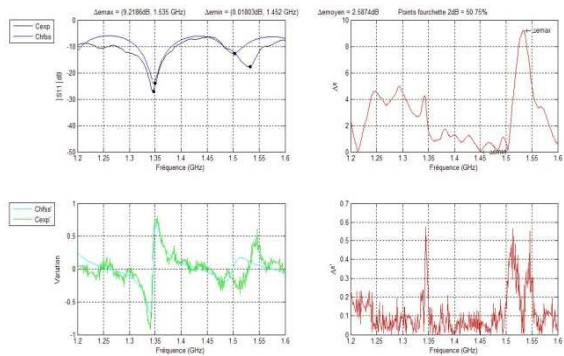

**Figure 4 :** *Exemple de résultats*

Le deuxième critère est le nombre de points compris dans une fourchette d'erreur autour des valeurs expérimentales: Cette fourchette d'erreur est exprimée en dB. Le programme calcule le nombre de points présents dans cette fourchette, exprimé en %. La meilleure courbe est celle qui maximise cette valeur ce qui symbolise un faible écart entre les courbes considérées.

Le troisième critère est le calcul de l'erreur moyenne entre la courbe expérimentale et les courbes simulées. La meilleure courbe est bien sûr celle qui minimise cette valeur.

Enfin, le dernier critère est le calcul de l'erreur maximale entre la courbe expérimentale et les courbes simulées. La meilleure courbe est celle qui minimise cette valeur.

Le nombre important de paramètres à manipuler et de résultats à visualiser nous ont conduits à développer une interface graphique d'utilisation. Cette dernière est appelée GUI (Graphical User Interfaces) dans Matlab. Les dernières versions de Matlab (ultérieures à 5.0-1997) disposent d'un outil baptisé GUIDE (Graphical User Interface Development Environment) spécifique à la création de ce type d'interface.

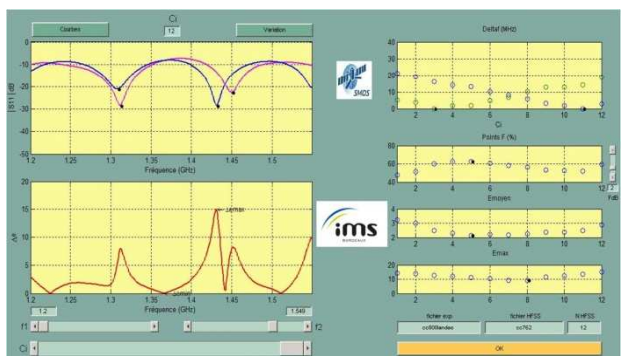

**Figure 5 :** *Interface utilisateur*

Nous avons donc créé une interface utilisateur afin de simplifier l'utilisation du logiciel (figure 5). Dans la partie supérieure gauche, la courbe expérimentale est représentée avec les courbes simulées. Des points noirs indiquent les pics de résonance détectés. Dans la partie inférieure gauche, c'est la courbe d'erreur qui est représentée. La courbe simulée représentée est sélectionnée avec le *slider* Ci, dans la partie basse. On peut également visualiser les taux de variation, en appuyant sur le bouton nommé "variation". Ce taux de variation est simplement calculé comme la différence entre deux échantillons du signal, et constitue une approximation de la dérivée. Les *sliders* "f1" et "f2" définissent la plage de fréquences représentée et utilisée pour réaliser les comparaisons.

Dans la partie de droite, les courbes des différents critères sont représentées. De haut en bas nous trouvons les valeurs pour les décalages des pics, les points dans la fourchette d'erreur, l'erreur moyenne et l'erreur maximale pour chaque courbe simulée. Le *slider* "FdB" permet à l'utilisateur de changer la taille de la fourchette d'erreur. Les noms de fichiers et le nombre de fichiers HFSS sont introduits dans les espaces indiqués, dans la partie inférieure à droite. Une analyse automatique et successive d'un ensemble de fichiers est possible ce qui permet un traitement rapide des données.

Dans cette première implémentation du système d'exploitation, c'est l'utilisateur qui doit faire la correspondance entre les courbes simulées Ci et les caractéristiques électromagnétiques de la structure représentée. Les résultats des critères sont présentés en fonction des Ci ; l'axe des abscisses indiquant le numéro de la courbe Ci. Il faut noter que cette implémentation n'est pas adaptée à l'utilisation d'une base de données importante. Dans des futures versions du système, un système d'entêtes des fichiers HFSS sera utilisé. Il est en cours de mise au point et il permettra une gestion de la future base de données de courbes simulées. Cela permettra au système d'exploitation de montrer sur l'écran les caractéristiques électromagnétiques des meilleures courbes, améliorant ainsi la démarche d'exploitation.

## 4. Résultats

Le cas pratique que nous présentons a été réalisé dans la forêt des landes (Site du Bray 33 France). Cette étude a été réalisée dans le cadre d'une campagne de mesures destinée à valider l'algorithme de traitement des données de la mission spatiale SMOS (Soil Moisture and Ocean Salinity) [4] [5] [6].

La figure 7 montre que l'erreur commise entre les fréquences 1.34 GHz et 1.45 GHz est de 1dB au maximum. Les courbes sont presque identiques, l'identification est donc possible. Nous pouvons attribuer les différences observées (beaucoup plus grande que les écarts trouvés lors d'études préliminaires en laboratoire) à plusieurs facteurs. D'abord, nous devons prendre en compte les erreurs de mesure expérimentale, c'est à dire, les erreurs de mesure de la hauteur, de l'inclinaison de(s) l'antenne(s) et les erreurs de mesure de l'analyseur. Nous devons aussi prendre en compte les erreurs de modélisation (taille du domaine de calcul, raffinement du maillage, critères de convergence du calcul …).

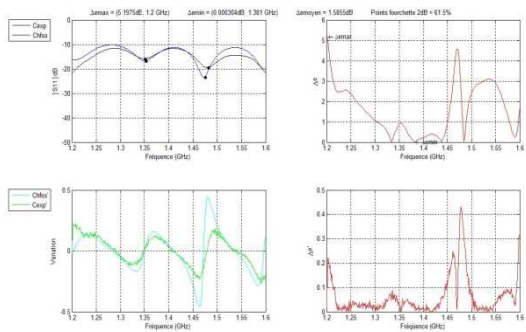

**Figure 6 :** *Exploitation sur un cas réel 1.3 GHz – 1.5 GHz*

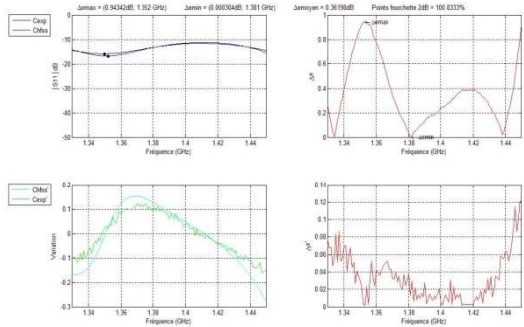

**Figure 7 :** *Exploitation sur un cas réel 1.34 GHz – 1.45 GHz*

De plus, la surface du terrain n'est pas parfaitement plane (des rugosités sont présentes), et le milieu n'est pas homogène (inclusions, gradient d'humidité...). Le modèle numérique utilisé pour créer la base de données simulées exploitée lors de cette étude ne prenait pas encore en compte la rugosité [3].

En utilisant des mesures de permittivité de la terre en fonction de l'humidité obtenues au laboratoire IMS [7] [8], il est possible, à partir de la permittivité retrouvée pour la terre, d'estimer l'humidité du sol mesuré. Les valeurs de la partie réelle et imaginaire de la permittivité obtenue, ε=6-j0.1, sont cohérentes et correspondent à une humidité de la terre aux alentours de SM = 17% ; valeur mesurée in-situ à l'aide de capteurs.

## 5. Conclusion et perspectives

Les résultats des modélisations ont montré qu'il est possible d'obtenir des courbes simulées similaires aux courbes expérimentales, mais que la modélisation des milieux géologiques n'est pas triviale. Si nous souhaitons obtenir de résultats plus satisfaisants, des calibrations du domaine d'intégration numérique plus précises se montrent nécessaires, ainsi qu'une modélisation plus complexe des milieux géologiques, en incluant des rugosités et des sources d'hétérogénéité par exemples (gradient d'humidité...). Toutefois, les résultats sont encourageants. Le modèle testé dans cette étude fournit des valeurs de permittivités cohérentes avec d'autres mesures pour les milieux caractérisés.

L'utilisation du système d'exploitation lors des modélisations effectuées a démontré son utilité pour faciliter et améliorer l'interprétation des résultats. Les critères implémentés se sont montrés performants en permettant de choisir les meilleures courbes. Les perspectives de développement du système passent par la mise en place d'un système d'entêtes pour les fichiers simulés HFSS. Cela rendra possible une obtention plus directe des résultats. De nouveaux critères pourront être aussi définis et testés. Ils complèteront les critères existants et amélioreront les performances globales du système d'exploitation.

Notre outil va être utilisé pour l'exploitation de données déjà recueillies (site du Pyla 33 France) et lors de nouvelles mesures comme par exemple le suivi en continu de propriétés de sols.

## Références

[1]. Ansoft website: http://www.ansoft.com.

[2]. Demontoux F., Ruffié G. and Wigneron J.P, "*Amélioration de l'étude de l'humidité de sols par radiométrie. Caractérisation et modélisation diélectriques de profils géologique*"' Journées Nationales Microondes, Nantes, May 2005.

[3]. F. Demontoux, H. Lawrence, JP Wigneron, "*Modélisation des effets de la rugosité sur l'étude de l'humidité des sols par radiométrie micro-ondes. Application à la mission spatiale SMOS*" jfmma telecom 08 Agadir Maroc

[4]. ESA website:http://www.esa.int/esaLP/LPsmos.html.

[5]. CNES website: http://www.cnes.fr/web/821-smos.php.

[6]. EADS Space website:http://www.space.eads.net/

[7] B. Le Crom, F. Demontoux, G. Ruffie, JP. Wigneron, JP. Grant *"Caractérisation électromagnétique de matériaux géologiques en vue du suivi de l'humidité des sols par radiométrie micro-ondes"* Jfmma & Telecom 2007 – Fes, Maroc

[8] F. Demontoux, B. Le Crom, G. Ruffié, J.P. Wigneron, J.P. Grant, V.L. Mironov, and H. Lawrence. *"Electromagnetic characterization of soil-litter media-Application to the simulation of the microwave emissivity of the ground surface in forests"* Eur. Phys. J. Appl. Phys. (2008 à paraître)